\documentclass[12pt,preprint]{aastex}

\usepackage{epsfig}

\usepackage{xcolor}

\usepackage{amsmath,amssymb,gensymb}

\begin{document}

%

\title{Systematic Redshift of the Fe III UV Lines in Quasars.  Measuring Supermassive Black Hole Masses under the Gravitational Redshift Hypothesis.}


%
%


%
%
%

%

\author{E. MEDIAVILLA\altaffilmark{1,2}, J. JIM\'ENEZ-VICENTE\altaffilmark{3,4}, C. FIAN\altaffilmark{1,2}, J. A. MU\~NOZ\altaffilmark{5,6}, E. FALCO\altaffilmark{7}, V. MOTTA\altaffilmark{8} \& E. GUERRAS\altaffilmark{9}}

\altaffiltext{1}{Instituto de Astrof\'{\i}sica de Canarias, V\'{\i}a L\'actea S/N, La Laguna 38200, Tenerife, Spain}
\altaffiltext{2}{Departamento de Astrof\'{\i}sica, Universidad de la Laguna, La Laguna 38200, Tenerife, Spain}
\altaffiltext{3}{Departamento de F\'{\i}sica Te\'orica y del Cosmos, Universidad de Granada, Campus de Fuentenueva, 18071 Granada, Spain}
\altaffiltext{4}{Instituto Carlos I de F\'{\i}sica Te\'orica y Computacional, Universidad de Granada, 18071 Granada, Spain}
\altaffiltext{5}{Departamento de Astronom\'{\i}a y Astrof\'{\i}sica, Universidad de Valencia, 46100 Burjassot, Valencia, Spain.}
\altaffiltext{6}{Observatorio Astron\'omico, Universidad de Valencia, E-46980 Paterna, Valencia, Spain}        
\altaffiltext{7}{Harvard-Smithsonian Center for Astrophysics, 60 Garden St., Cambridge, MA 02138, USA}
\altaffiltext{8}{Instituto de F\'{\i}sica y Astronom\'{\i}a, Facultad de Ciencias, Universidad de Valpara\'{\i}so, Avda. Gran Breta\~na 1111, 2360102 Valpara\'{\i}so, Chile}
\altaffiltext{9}{Homer L. Dodge Department of Physics and Astronomy, The University of Oklahoma, Norman, OK, 73019, USA}

\begin{abstract}
We find that the Fe III$\lambda\lambda$2039-2113 spectral feature in quasars appears systematically redshifted by amounts  {accountable}  {under the hypothesis of} gravitational
 redshift induced by the central supermassive black hole. {Our} analysis of  27 composite spectra from the BOSS survey {indicates} that the redshift and the broadening of the lines in the Fe III$\lambda\lambda$2039-2113 blend {roughly follow the expected  correlation} in the weak limit of Schwarzschild geometry  {for virialized kinematics}. {Assuming that} the Fe III UV redshift {provides} a measure of $M_{BH}\over R$ (${\Delta \lambda\over \lambda}\simeq{3\over2}{G\over c^2} {M_{BH}\over R}$) and  {using different} estimates {of the emitting region size, $R$} (either from  {gravitational microlensing,} reverberation mapping or from the scaling of size with intrinsic quasar luminosity), we obtain masses for 10 objects which are in agreement within uncertainties with previous mass estimates based on the virial theorem. 
Reverberation mapping estimates of the size of the  Fe III$\lambda\lambda$2039-2113 emitting region in a sample of objects  would be needed to confirm the gravitational origin of the measured redshifts.
Meanwhile, we present a {tentative} black hole mass scaling relationship based on the Fe III$\lambda\lambda$2039-2113 redshift {useful to measure the black hole mass of one individual object from a single spectrum}.


\end{abstract}


\section{Introduction \label{intro}}

In the classical picture of quasars, a central supermassive black hole (BH) is surrounded by an inspiraling disk  that transports matter into the depth of the gravitational well of the BH, releasing huge quantities of energy (Zeldovich 1964, Salpeter 1964). This central engine illuminates gas clouds located in a larger region (Broad Line Region, BLR) giving rise to very broad emission lines (BEL) whose width and shape are determined by the kinematics of the gas clouds, ultimately ruled by the central BH. Thus, the kinematics of the BLR potentially provides a means of measuring the central masses of supermassive BH and of studying the structure of the accretion disk.

Specifically, the methods for estimating BH masses in distant quasars\footnote{In the nearby universe, masses of supermassive black holes have been determined in around 70 galaxies by direct modeling of the stellar or gas dynamics (see, e.g., McConnell \& Ma 2013).} are  mainly based on the measure of the broadening of the BEL in combination with the virial theorem (see, e.g., Peterson 2014). According to this theorem, the square of the line-broadening, $(\Delta v)^2$, is a proxy for $M/R$ that, in combination with a determination of the size, $R$, can provide an estimate of the mass,

\begin{equation}
\label{virial}
M=f{(\Delta v)^2R\over G}.
\end{equation}
The dimensionless factor, $f$, includes the effects of the unknown BLR geometry, kinematics and inclination.  {Without} more information, it is a common practice to use an average value for $f$ obtained {by} calibrating with other methods\footnote{The $M_{BH}-\sigma_*^2$ relationship, for instance (Ferrarese \& Merritt 2000, Gebhardt et al. 2000, Tremaine et al. 2002).}, even when $f$ is different for each object. This virial factor, by itself, limits  the accuracy of individual estimates of mass to $\sim$0.4 dex (Peterson 2014). The size can be determined from reverberation mapping (see, e.g., the reviews by Peterson 1993, 2006), which is an observationally expensive technique, or alternatively using the size-luminosity, R-L, relationship for AGN, a shortcut inferred from reverberation mapping results (Kaspi et al. 2000, 2005, Bentz et al. 2009, Zu et al. 2011). Both techniques are relatively accurate and the main experimental problem  to apply  Eq. \ref{virial} (in addition to the unknown factor $f$),  arises from the determination of the line widths (Peterson 2014), due to both,  the ambiguity in the definition of $\Delta v$ (FWHM, $\sigma$, use of the variable or constant part of the spectra, etc.), and the presence of contaminating features (extra components, blended lines, pseudo-continuum, etc.).

An alternative path to BH masses is the gravitational redshift of the BEL. If we consider the width of the BEL as caused by motion in the gravitational field of a central mass, a simple calculation shows that we should expect measurable gravitational and transverse Doppler redshifts (see, e.g., Netzer 1977, Anderson 1981, Mediavilla \& Insertis 1989). Indeed, 
in the weak limit of the Schwarzschild metric, the velocity of the emitters is proportional to $\sqrt{GM/R}$, and  the gravitational plus transverse Doppler redshift\footnote{Hereafter we refer to the combined  gravitational and transverse Doppler effects as gravitational redshift.}  will tend to ${3\over 2}{G\over c^2}{M\over R}$. Thus, line broadenings typical of the BEL, $\Delta v \gtrsim 10^3\rm\, km\, s^{-1}$, will result in redshifts $z_{grav}={\Delta \lambda \over \lambda} \gtrsim 0.003$, which for UV lines corresponds to displacements $\Delta \lambda \gtrsim 6$\AA, which should be very easy to measure.

However, experimental results do not satisfy these theoretical expectations. According to massive analysis of quasar spectra like SDSS (Vanden Berk et al. 2001) and BOSS (Harris et al. 2016),  the peaks of the brightest permitted and semi forbidden lines can appear shifted either towards the red or the blue, but with blue shifts being more frequent and strong, the opposite to what is expected. This indicates that the shifts of the BEL peaks are probably of kinematic origin. Nevertheless, the profiles of some BEL (H$\beta$ in many cases) can show redward asymmetries (Peterson et al. 1985, Sulentic 1989, Zheng \& Sulentic 1990, Popovi{\'c} et al. 1995, Corbin et al. 1997), that have been sometimes interpreted as the result of gravitational redshift (Joni\'c et al. 2016), although  the presence of an extra component redshifted due to inflow is, perhaps, a more accepted explanation for the line asymmetry. In a few cases in which spectroscopic monitoring is available, the redshift between the mean and rms profiles of Balmer lines has been associated {with} gravitational redshift (Kollatschny 2003, Liu et al. 2017)\footnote{In any case, the presence of gas cold enough as to generate the Balmer lines so close to the BH as to justify the redshift needs to be explained (Bon et al. 2015).}. There has also been continued controversy about the existence of redshifts in Fe II emission. {Hu et al. (2008) interpreted the redshift measured in the Fe II optical lines of a sample of SDSS spectra in terms of kinematics dominated by infall. In this scenario, to prevent the gas from being accelerated away from the central source by the radiation force, Ferland et al. (2009) propose that we only observe the shielded face of near-side infalling clouds. However, Kova{\v c}evi{\'c} et al. (2010) report only a slight redshift of the Fe II optical lines. The same result is reached by Sulentic et al. (2012) who find that the Fe II optical lines follow the same kinematics as the Balmer lines. Finally, Kova{\v c}evi{\'c}-Doj{\v c}inovi{\'c} \& Popovi{\'c} (2015) find a significant average redshift in the UV lines that, however, is not present in the optical lines}. In any case, {the Fe II redshifts} were interpreted as inflow of gas clouds located at the outer parts of the BLR, leaving aside the gravitational redshift scenario.

The main cause of  the scarcity of unquestionable identifications of gravitational redshift is likely the complex morphology of the lines, with several kinematic components arising from different regions, and often blended with lines from other species that may significantly distort the shape and change the width of the line profile. To achieve a robust detection of gravitational redshift we need a feature associated {with} one single ion, not blended  with emission lines of other species, and that presumably originates from an inner region of the BLR.

The size of the region giving rise to an emission line in the quasar spectrum can be estimated from the changes in magnification of the emission line induced by gravitational microlensing\footnote{When a distant quasar is lensed by the gravitational potential of an intervening (lens) galaxy, the relative movement between the quasar and the distribution of stars in the lens galaxy can change the brightnesses of the images, an effect called quasar gravitational microlensing (Chang \& Redfsdal 1979, 1984, see also the review by Wambsganss 2006).}, so that the larger the changes the smaller the size. According to previous studies (Guerras et al. 2013a,b, {Fian et al. 2018}), the Fe III$\lambda\lambda$2039-2113 blend is a relatively isolated feature strongly affected by microlensing and hence must {originate} in a small region (a few light-days across) where gravitational redshift is significant. The objective of this {work is, then,}  to measure the shifts of the Fe III lines of this blend, to {explore} their consistence with the gravitational redshift hypothesis, and to discuss their possible use  in the determination of SMBH masses and in the study of the physics of accretion disks.

{The paper is organized as follows: In \S 2 we fit the Fe III$\lambda\lambda$2039-2113 blend in a sample of high S/N spectra collected from several data sources. \S 3 is devoted to deriving a scaling relationship of mass with redshift and luminosity. Finally, in \S 4 we summarize the main conclusions.}

\section{Results: Fe III$\lambda\lambda$2039-2113 Redshift Measurements\label{proxy}}

\subsection{Data}
{The data analyzed in this work have different origins. The 14 lensed quasar spectra fitted in  \S \ref{results} have been compiled from many sources in the literature (see details in Fian et al. 2018). In \S \ref{results} we also analyze the publicly available SDSS composite spectrum (Van den Berk et al. 2001) and the 27 BOSS quasar composite spectra (Jensen et al. 2016). Finally, in \S \ref{RM} the monitoring series of spectra of NGC 5548 (Korista et al. 1995) and the spectrum from NGC 7469 (Kriss et al. 2000) are used to determine the BH masses. All the spectra are corrected from cosmological redshift.}

%
\subsection{Analysis and results\label{results}}

\begin{figure}[h]
\vskip -3 truecm
\includegraphics[scale=0.75]{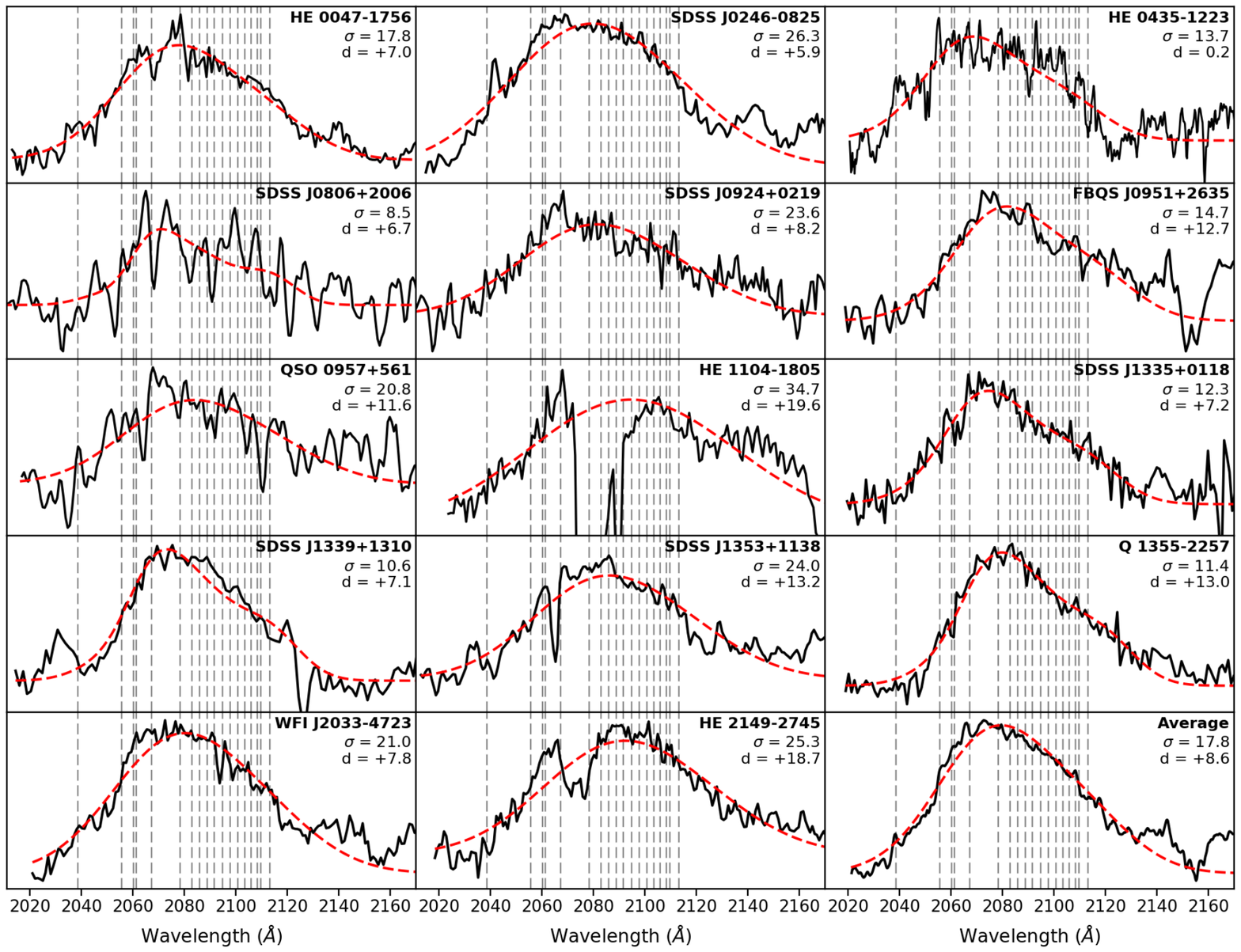}
\vskip -1 truecm
\caption{Fits to the Fe III$\lambda\lambda$2039-2113 blend in 14 lensed quasars. The broadening, $\sigma$, and shift, $d=\Delta \lambda$, of the iron lines are indicated for each spectrum (in \AA). The continuous (dashed) curve corresponds to the data (fit). Vertical dashed lines are located at the wavelengths corresponding to the Fe III lines of {the} Vestergaard \& Wilkes (2001) template at rest.  The spectra have been shifted by an amount $-d$ to match the template rest frame.\label{lenses}}
\end{figure}

\begin{figure}[h]
\includegraphics[scale=1.3]{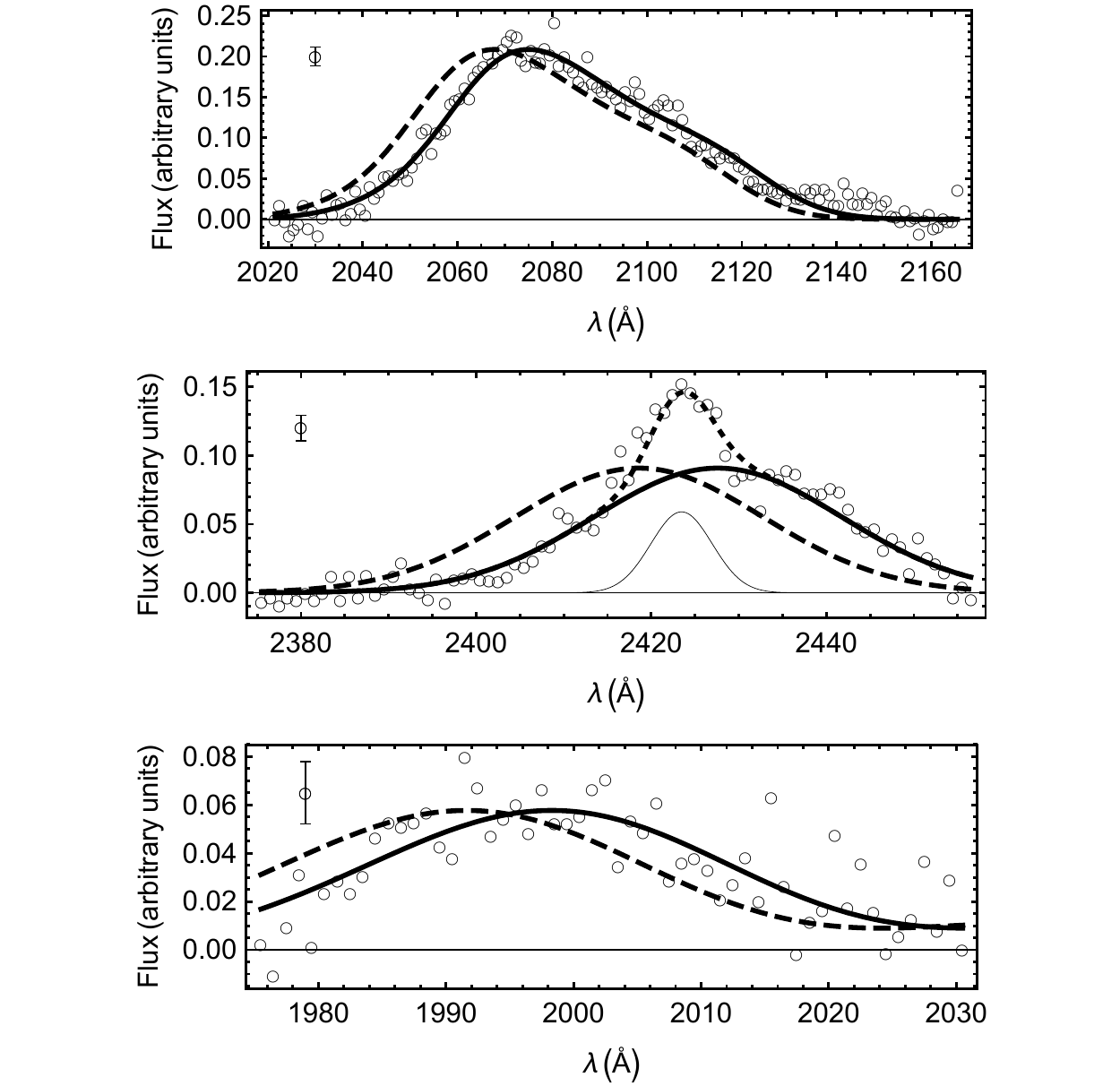}
\caption{Fits of three UV iron features, Fe III$\lambda\lambda$2039-2113 (upper panel), Fe III$\lambda\lambda2419$  (middle panel) and  Fe III$\lambda\lambda$1970-2039  (lower panel), in the SDSS composite spectrum (Van den Berk et al. 2001). Notice the redshifts between the rest frame features (dashed curves) and the best fits (solid curves). In the case of   Fe III$\lambda2419$ (middle panel) the thin curve represents the unshifted, narrow Ne IV$\lambda2424$ line, and the dotted curve the total fit. Average error bars are included in each panel.\label{fit_SDSS}}
\end{figure}

We  model the Fe III$\lambda\lambda$2039-2113 {spectral}  feature in {14} lensed quasars ({Fian et al., 2018). First, we fit the continuum to a straight line defined in two windows at the blue ($2013.3\rm\,\AA,2017.9\rm\,\AA$) and red ($2195.3\rm\,\AA,2205.0\rm\,\AA$)  sides of the blend. Then we subtract the continuum and}  {fit the feature using a template of 19 single Fe III lines between 2038.5 and 2113.2\AA\  of fixed relative amplitudes as provided by Vestergaard \& Wilkes (2001). The (Gaussian) lines are broadened, shifted and scaled with the same width, $\sigma\ (=\rm FWHM/2.35)$,  wavelength shift, $\Delta \lambda$, and scale factor.  In Figure \ref{lenses} we can see that this template is able to reproduce very well the shape of the Fe III$\lambda\lambda$2039-2113 feature in the spectra of the objects in our sample\footnote{See also other fits in the upper panel of Figure \ref{fit_SDSS}, and Figures \ref{grid} and \ref{grid2}.}, but the fitted features are redshifted in all the objects {except one (SDSS 1004+4112, which is strongly affected by microlensing). Leaving aside this object, we find that the average of this systematic redshift is $\langle \Delta \lambda \rangle=10.3\rm\, \AA$ with a scatter between objects of $\pm 5.9\rm\, \AA$. If we take the microlensing based size inferred by Fian et al. (2018) for the Fe III UV lines\footnote{Fiann et al. (2018) consider a disk with a Gaussian radial profile, for which the half-light radius, $R$, is obtained from the reported Gaussian sigma, $r_s$, through $R=1.18 r_s$.}, $R=1.18\times 11.3^{+5}_{-4}$ light-days, we can estimate the average mass of the supermassive black holes of the lensed quasars under the hypothesis of a gravitational origin for the redshift. {If we assume that gravitational and transverse Doppler are the physical phenomena giving rise to the redshift, we have (see, e.g.,  Mediavilla \& Insertis, 1989), $\nu=(\nu_0/\gamma)\sqrt{1-2GM_{BH}/Rc^2}$ with $\gamma=\left(1-(v/c)^2\right)^{-1/2}$}. In the weak limit of the Schwarzschild metric, $v\simeq GM_{BH}/R$, and we have, 

\begin{equation}
\label{redshift}
z_{grav}={\Delta \lambda\over \lambda}\simeq{3\over2}{G\over c^2} {M_{BH}\over R},
\end{equation}
and,
\begin{equation}
\label{mass}
M_{BH}\simeq{2 c^2\over3 G}{\Delta \lambda\over \lambda}{R}=\left({z_{grav} \over 0.005}\right) \left({R\over 10 {\rm \, light\, days}}\right) \left(0.58 \times  10^9 M_\odot\right).
\end{equation}
Susbtituting in Equation \ref{mass} the mean redshift of the iron lines and the microlensing based size, we obtain for the average mass of the supermassive black holes, $\langle M_{BH}\rangle \simeq (0.83\pm 0.47)\times 10^9 M_\odot$, where the uncertainty arises partly from the method and partly from the intrinsic scatter between objects. This value is in good agreement, in mean and scatter, with virial based estimates for lensed quasars (see, e.g., Figure 8 of Mosquera et al. 2013). In fact, if we consider the 8 lensed quasars in our sample (HE 0047-1756, SDSS 0246-0285, SDSS 0924+0219, FBQ 0951+2635, Q 0957+561, HE 1104-1805, SDSS 1335+0118 and HE 2149-2745) that have virial mass estimates by Peng et al. (2006) and Assef et al.  (2011), we obtain from Eq. \ref{mass}, $\langle M^{micro}_{BH}\rangle \simeq (0.9\pm 0.5)\times 10^9 M_\odot$, in very good agreement with the average of their virial masses, $\langle M^{virial}_{BH}\rangle \simeq 0.93\times 10^9 M_\odot$.

}}

{Because of the interesting implications of {these results}, and to exclude any systematic issue in our sample of lensed quasars, we fit the Fe III$\lambda\lambda$2039-2113 feature in the high S/N composite SDSS spectrum (Vanden Berk et al. 2001), in which we also measure a strong global redshift of the feature of $\sim 7\,\rm\AA$ (Figure \ref{fit_SDSS}).  Looking for further confirmation,  we fit another two UV features of Fe III that, in spite of their lower intensity, can be modeled in this high S/N composite spectrum: the Fe III$\lambda$2419 line and the Fe III$\lambda\lambda$1970-2039 blend. The Fe III$\lambda$2419 line (Figure \ref{fit_SDSS}) appears blended with a narrow line identified as Ne IV$\lambda2424$ (Vanden Berk et al. 2001). Figure  \ref{fit_SDSS} shows that, while the Ne IV narrow line can be well fitted at its nominal wavelength (Vanden Berk et al. 2001), the Fe III line has a clear redshift with respect to it. Finally, the redshift is also observed in the (noisier) Fe III$\lambda\lambda$1970-2039 blend. 
The best fit estimates of the redshift, $z=\Delta \lambda / \lambda$, of these features are: $0.0034\pm0.0002$ (Fe III$\lambda\lambda$2039-2113), $0.0037\pm0.0001$ (Fe III$\lambda$2419) and $0.0034\pm 0.0007$ (Fe III$\lambda\lambda$1970-2039). For the widths, $\sigma / \lambda$, we obtain: $0.0057\pm0.0003$ (Fe III$\lambda\lambda$2039-2113), $0.0059\pm0.0002$ (Fe III$\lambda$2419) and $0.0055\pm 0.0006$ (Fe III$\lambda\lambda$1970-2039). 
The good agreement between the fitted parameters of the three Fe III features confirms that the redshift is intrinsic to the Fe III emitters.}

Going a step further, to study the incidence and meaning of the observed redshift using high S/N spectra, we fit  {(see Figures \ref{grid} and \ref{grid2})} the Fe III$\lambda\lambda$2039-2113 feature in the 27 composite spectra of the BOSS survey (Jensen et al. 2016). {The fits are very good with $\chi_{red}^2 \le 2$, although some of the spectra have a low S/N ratio. {We can use BOSS composites to discuss virialization. If the kinematics is virialized (Eq. \ref{virial}), we should have,

\begin{equation}
\label{virialsigma}
G{M_{BH}\over R}= f (\Delta v)^2 = f \left(\sigma \over \lambda\right)^2c^2,
\end{equation}
where we have taken $ \sigma c / \lambda$ as representative of the line broadening\footnote{For our Gaussian based fits, $\sigma=FWHM/2.35$ but in many applications of the virial theorem based on emission-line profiles, $\sigma$ is the second moment of the experimental line profile, and $FWHM/\sigma$ depends on the profile shape (Collin et al. 2006).}, $\Delta v$. Combining Eq. \ref{virialsigma} with the expression for the mass in terms of the redshift (Eq. \ref{redshift}),  we obtain, 

\begin{equation}
\label{f}
{\Delta \lambda\over \lambda} ={3\over 2} f \left(\sigma \over \lambda\right)^2.
\end{equation}
Taking logarithms, we can write this condition of virialized kinematics in a linear shape convenient for quantitative fitting, 

\begin{equation}
\label{flog}
\log \left({\sigma \over \lambda}\right)^2=-\log{3 f\over 2}+ \log\left(\Delta \lambda \over \lambda\right).
\end{equation}
The measured redshifts, $z=\Delta \lambda/\lambda$, and widths of the Fe III lines, $\left(\sigma \over \lambda\right)^2$, obtained from the BOSS composite spectra (excluding the cases with S/N $< 3.0$) follow this correlation though with a relatively high scatter (Figure \ref{lambda_vs_sigma2}). Fitting Eq. \ref{flog} to the data we obtain {(R-squared $\sim 0.75$)},

\begin{equation}
\label{log}
\log \left({\sigma \over \lambda}\right)^2=-2.09\pm0.64+ (0.99\pm0.26)\log\left(\Delta \lambda \over \lambda\right).
\end{equation}

}

{The large uncertainties in the fit parameters (Eq.  \ref{log}) can have an intrinsic origin, for the virial factors, $f$, can be significantly different from system to system depending on  physical unknowns like the flatness of the emitter's distribution, its orientation, or the presence of non gravitational forces (e.g., radiation pressure). It is likely that the criteria to form the BOSS composites may be biased with respect to any of these unknowns giving rise to an intrinsic scatter in $f$. On the other hand, radial motions may also contribute to the redshift  in a variable way from object to object, increasing the scatter. In any case, } {alternative explanations {(inflow, for instance, may be another mechanism giving rise to the redshifts)} would need additional physics to explain {the observed trend between broadening and redshift. Thus, while a tight correlation between $\Delta \lambda/\lambda$ and  $\left(\sigma \over \lambda\right)^2$ is not generally expected, the trend found between these two quantities among the composite spectra of BOSS supports the gravitational interpretation of the  Fe III$\lambda\lambda$2039-2113 redshifts and indicates that the kinematics is not far from virialized.}}

 {Although the fits of the Vestergaard \& Wilkes (2001) template to the Fe III$\lambda\lambda$2039-2113 feature of BOSS composites are very good, it is true that this template is based on one particular object. To eliminate any possible bias related to the use of the template, we have performed an alternative study based on the centroid of the blend, $\lambda_c=\langle \lambda \rangle$, in each composite spectrum. The standard deviation between the redshift measurements based on either the fit of the template or the centroid of the blend is $\sim0.5\rm\,\AA$.  This result confirms the redshift estimates irrespective of the choice of template. Another possible source of uncertainty in the measurement of the redshifts is the difficulty to determine the systemic velocity of the quasars, which may depend on the choice of the spectral features. However, this indetermination can account for shifts of roughly a few hundred $\rm km\, s^{-1}$, randomly distributed between blue- and red-shifts while we are measuring exclusively redshifts of about one thousand $\rm km\, s^{-1}$. In addition, this problem should be mitigated in the case of BOSS composites resulting from the average of many spectra.}

}

{\section{Discussion: Black Hole Mass Estimates Based on Fe III$\lambda\lambda$2039-2113 Redshift}

Under the hypothesis that the redshift of the Fe III$\lambda\lambda$2039-2113 is of gravitational origin, we can invert Equation \ref{redshift} to derive the central BH mass corresponding to any object for which an estimate of $R_{FeIII}$ can be obtained (see Eq. \ref{mass}). We are going to consider three different methods for computing sizes: reverberation mapping, scaling of the size of the BLR with luminosity and gravitational microlensing.}

\subsection{Mass Estimates of the Central Black Holes in NGC 5548 and NGC 7469 based on Fe III$\lambda\lambda$2039-2113 Redshift and Reverberation Mapping \label{RM}}

{NGC 5548 is a widely studied AGN\footnote{Notice, however, that some common conceptions about this AGN could change if the suspected existence of a supermassive BH binary in the center of this galaxy (Li et al. 2016) is confirmed.} for which reverberation mapping has yielded estimates of the size for the continuum and several strong emission lines (see, e.g., Clavel et al. 1991, Korista et al. 1995, Peterson et al. 2002; see also Pei et al. 2017 and references therein).

We fit the Fe III$\lambda\lambda$2039-2113 blend in each of the spectra of the monitoring series (Korista et al. 1995), deriving the light curve of the Fe III amplitude (Figure \ref{NGC5548}). We infer a lag of the Fe III relative to the UV$\lambda 1970$ continuum of $3.3\pm0.8$ ($2.8\pm1.4$) days when the centroid (peak) of the cross correlation centroid (peak) distribution CCCD (CCPD) is taken as reference.} The errors have been estimated applying flux randomization Monte Carlo methods. {Adopting these lags as estimates of $R_{FeIII}$ and using the measurement of the redshift from the fit to the average spectra, $z_{grav}(Fe III)=(\Delta \lambda / \lambda)_{Fe III} =0.0056\pm 0.0010$, we  obtain, $M_{BH}=2.2^{+0.6}_{-0.4}\times 10^8M_\odot$ ($M_{BH}=1.8^{+1.0}_{-0.9}\times 10^8M_\odot$) for the centroid (peak). These values are relatively large but in agreement within uncertainties with recent estimates of the black hole mass derived from the virial {theorem} ($M=1.2^{+0.4}_{-0.3} \times 10^8M_\odot$, Ho \& Kim, 2015; $M=6.7^{+2.7}_{-2.7} \times 10^7M_\odot$, Pei et al 2017), taking into account a 30\% uncertainty in the average virial factor $f$  (Woo et al. 2015), and the intrinsic scatter between objects (0.35 dex according to Ho \& Kim 2015). 

We also fit the  Fe III$\lambda\lambda$2039-2113 feature in another well studied galaxy, NGC 7469 (Kriss et al. 2000). We measure $z_{grav}(Fe III)=0.0026\pm 0.0005$. In this case there is no UV spectroscopic monitoring to obtain the light curve of the Fe III blend, but we can set an upper limit to the size of $\sim0.7$ light-days. This value corresponds to the reverberation lag of He II. This is a high ionization line, known from the impact of microlensing (Fian et al. 2018) to arise from a region of size comparable or somewhat greater than that corresponding to Fe III. Taking this upper limit, we infer  $M_{BH}\leq2.1^{+0.4}_{-0.4}\times 10^7M_\odot$, compatible with {previous} virial estimates ($M =1.5^{+0.6}_{-0.4}\times 10^7M_\odot$, Ho \& Kim, 2015, $1-6\times 10^7M_\odot$, {Shapovalova et al. 2017}).
}


Finally, it is also important to stress that, once the size is known via reverberation mapping, the mass of the object is directly obtained from the redshift without using any previous calibration, i.e., in combination with reverberation mapping, the gravitational redshfit of the Fe III$\lambda\lambda$2039-2113 feature is a {\sl primary} method to determine masses. In fact, because gravitational redshift does not depend on geometrical considerations, it may become the primary calibrator of all the other methods used to measure the mass of the BH.

\subsection{Black Hole Mass Estimates Based on Fe III$\lambda\lambda$2039-2113 Redshift and Quasar Luminosity.\label{BOSS}}

{
Reverberation mapping is an observationally expensive technique to estimate sizes. An alternative  is to use the scaling of the size of the BLR with luminosity, {$R\propto  \left(\lambda L_\lambda\right)^\alpha$} (Kaspi et al. 2000, 2005){. In combination with the line width of the BLR lines as an estimator of the virial velocity,} empirical BH mass calibrations, $M_{BH}\propto FWHM^2 \left(\lambda L_\lambda\right)^\alpha${, can be obtained. The most reliable $R-L_\lambda$ relationship is based on H$\beta$ and $L_{5100}$. Other determinations, related to H$\alpha$, Mg II or CIV,  are re-calibrated from the $R(H\beta)-L_{5100}$ relationship. In spite of some problems associated with it (see, e.g., Mej\'\i a-Restrepo et al. 2016), the calibration using the CIV line is important because it is the only prominent broad emission line that lies within the optical window at high-z  as is the case in many of the objects we studied.}

Specifically, for high redshift quasars, BH masses can be estimated from the CIV$\lambda1549$ broadening using\footnote{The use of other standard calibrations (e.g. Vestergaard \& Peterson 2006, Assef et al. 2011) do not substantially affect the results.} {(Mej\'\i a-Restrepo et al. 2016)},

\begin{equation}
\label{VP06}
M_{BH}(CIV)=10^{6.353\pm 0.013}\left(FWHM_{CIV}\over 10^3\,{\rm km\,s^{-1}}\right)^2\left({\lambda L_\lambda (1450{\rm\,\AA})\over 10^{44} {\rm\, erg\,s^{-1}}}\right)^{0.599\pm 0.001}M_\odot.
\end{equation}
Thus, we can use the $FWHM_{CIV}$ measurements available for the BOSS composites (Jensen et al. 2016) to 
re-calibrate Eq. \ref{VP06} in terms of the Fe III gravitational redshift\footnote{This is supported by Eq.\ref{f} which relates broadenings and redshifts.}. 
On average, we find for the BOSS composite spectra: $ <FWHM_{CIV}>=(0.27\pm 0.02)<\sqrt{z_{grav}(Fe III)}>c$, where the uncertainty is the standard error in the mean. Substituting this in Eq. \ref{VP06} we obtain a mass scaling relationship based on the gravitational redshift of Fe III,

\begin{equation}
\label{atlast}
M^{BOSS}_{BH}(Fe III)=10^{7.69^{+0.06}_{-0.07}}\left(z_{grav}(Fe III)c\over 10^3\,{\rm km\,s^{-1}}\right)\left({\lambda L_\lambda (1350{\rm\,\AA})\over 10^{44} {\rm\, erg\,s^{-1}}}\right)^{0.599\pm0.001}M_\odot.
\end{equation}
To check the validity of this relationship, we compare in Figure \ref{mass_vs_mass_BOSS} the mass estimates obtained  applying Eq. \ref{atlast} to the measured Fe III gravitational redshifts of the lensed quasars in our sample (Fian et al. 2018) with the virial based masses obtained by Peng et al. (2006) and Assef et al. (2011). We have 8 objects in common: HE 0047-1756, SDSS 0246-0285, SDSS 0924+0219, FBQ 0951+2635, Q 0957+561, HE 1104-1805, SDSS 1335+0118 and HE 2149-2745. We have also included NGC 5548 and NGC 7469 in the plot (gravitational redshift masses obtained from Eq. \ref{atlast} and virial masses from Vestergaard \& Peterson 2006). The global agreement over two orders of magnitude in mass is very noticeable,  showing that the Fe III$\lambda\lambda$2039-2113 gravitational redshift can be used to measure the BH mass. 

The {intercept of the} best fit with slope unity (dashed line in Figure \ref{mass_vs_mass_BOSS}) corresponds to the {shift in the calibration} that we would obtain following the usual steps to derive the mass scaling relationships (see, e.g., Peterson et al., 2004; Vestergaard \& Peterson,  2006): (i) adopt a R-L relationship, $R\propto L^{0.599}$, and (ii) use the available virial based mass estimates to calibrate our unscaled masses, $\mu=\left(z_{grav}(Fe III)c/ 10^3\,{\rm km\,s^{-1}}\right) \left({\lambda L_\lambda (1350{\rm\,\AA})/ 10^{44} {\rm\, erg\,s^{-1}}}\right)^{0.599}M_\odot$
%
{. The relatively small value of the shift in the calibration, {0.04} dex, as compared with the $1\,\sigma$  scatter of the masses with respect to the best fit, {0.26} dex, indicates that there is a } good agreement between the BOSS composite spectra based calibration and the independent calibration that would be obtained fitting the virial masses. The  {0.26} dex scatter of the masses {relative} to the fit, also indicates that Eq. \ref{atlast} is reliable  taking into account that virial masses are themselves uncertain typically by $\sim 0.3\,\rm dex$ (Vestergaard \& Peterson 2006).

Notice that the R-L relationship is very tight (with errors comparable to the lags inferred from reverberation mapping, Peterson 2014). Thanks to this and because the Fe III gravitational redshift is easy to measure from a single spectrum, Equation \ref{atlast} provides a robust estimate of the mass of a quasar or AGN. An attempt to fit the Fe III$\lambda\lambda$2039-2113 blend in a sample of  $\sim$200 SDSS individual quasars (S/N $\gtrsim 20$) shows that the redshift can be measured with a reasonable accuracy in 25\% of them. This implies a number of potential BH mass determinations of more than one thousand from available and future quasar surveys.
}

%
%

{
\subsection{Black Hole Mass Estimates Based on Fe III$\lambda\lambda$2039-2113 Redshift and  Microlensing Size Scaling.\label{micro}}

Using microlensing based sizes, we can also estimate the BH masses directly from the equation of the redshift in the weak limit of the Schwarzschild metric (Eq. \ref{mass}). We do not have individual estimates of size for each object, but we can use the average microlensing size estimated by Fian et al. (2018)  re-scaling it by applying the $R\propto \sqrt{\lambda L_\lambda}$ relationship:

\begin{equation}
\label{massmicro}
M_{BH}\simeq{2 c^2\over3 G}{\Delta \lambda\over \lambda}\langle R\rangle{\sqrt{\lambda L_\lambda}\over \langle\sqrt{\lambda L_\lambda}\rangle} .
\end{equation}
This equation is, indeed, very similar to Equation \ref{atlast} but has been derived on different grounds.  Inserting the value of $\langle R\rangle$ from Fian et al. (2018) and the average of the square root of the luminosities of the quasars, $\langle\sqrt{\lambda L_\lambda}\rangle$, used by these authors to infer $\langle R\rangle$, we can write,

\begin{equation}
\label{massmicro2}
M_{BH}\simeq{}{\Delta \lambda\over 10.3{\,\rm \AA}}{\sqrt{\lambda L_\lambda\over 10^{45.79} {\rm\, erg\,s^{-1}}}}\times (0.83\pm 0.47)\times 10^9 M_\odot.
\end{equation}
It is convenient to rewrite this equation to compare it with the equivalent expression (Eq. \ref{atlast}) based on the BOSS composite spectra calibration,
\begin{equation}
\label{atlast2}
M^{micro}_{BH}(Fe III)=10^{7.85^{+0.20}_{-0.36}}\left(z_{grav}(Fe III)c\over 10^3\,{\rm km\,s^{-1}}\right)\left({\lambda L_\lambda (1350{\rm\,\AA})\over 10^{44} {\rm\, erg\,s^{-1}}}\right)^{0.5}M_\odot.
\end{equation}
Thus, Eqs. \ref{atlast} and \ref{atlast2} agree within uncertainties. This agreement is noteworthy taking into account that  the calibration of Eq. \ref{massmicro2} (and hence Eq. \ref{atlast2}) resides on gravitational microlensing while Eq. \ref{atlast} has been calibrated from the widths of the CIV lines of BOSS composites. Figure \ref{mass_vs_mass_micro} shows the good agreement, 0.27 dex of scatter ($1\,\sigma$), between the mass estimates obtained using Eq. \ref{massmicro2} and the virial masses. 

\subsection{Best Fit of the Mass Scaling Relationship to the Virial Masses Leaving Free the $R\propto L^b$ Law.\label{bestfit}}

Finally, it is also interesting to perform a fit of Equation \ref{massmicro} to the virial masses of our 10 objects but now leaving free the exponent of the R-L relationship, $R\propto \lambda L_\lambda^b$. A change of scale, $a$, is also allowed. Specifically we fit $a$ and $b$ parameters in,

\begin{equation}
\label{fitab}
\log{\left( {M_{vir}\over 0.83 \times 10^9 M_\odot}\right)}=\log{\left( a  {\Delta \lambda \over 10.3 {\rm\,\AA}}\right)} + b \log{\left({L_\lambda \over \langle \lambda L_\lambda^b\rangle^{1/b}}\right)},
\end{equation}
where $\langle \lambda L_\lambda^b\rangle$ is computed taking into account all the objects used by Fian et al. (2018) to estimate the average microlensing size. We obtain $a=1.1\pm0.3$ and $b=0.57\pm 0.08$. That is, Eq. (\ref{massmicro}) agrees within uncertainties with the best fit to virial masses. To show this explicitly we can, once more, write Eq. \ref{fitab} as,

\begin{equation}
\label{atlast3}
M^{best\, fit}_{BH}(Fe III)=10^{7.89^{+0.11}_{-0.13}}\left(z_{grav}(Fe III)c\over 10^3\,{\rm km\,s^{-1}}\right)\left({\lambda L_\lambda (1350{\rm\,\AA})\over 10^{44} {\rm\, erg\,s^{-1}}}\right)^{0.57\pm0.08}M_\odot,
\end{equation}
in agreement within uncertainties with both, $M^{micro}_{BH}$ (Eq. \ref{atlast2}) and $M^{BOSS}_{BH}$ (Eq \ref{atlast}). In Figure \ref{mass_vs_mass_bestfit}  we compare the results of $M^{bestfit}_{BH}$  with the virial masses ($1\,\sigma=0.26\rm\, dex$). 

Thus, according to the results discussed in sections \ref{BOSS}, \ref{micro} and \ref{bestfit}, from three different methods (Eqs. \ref{atlast}, \ref{atlast2} and \ref{atlast3}), we have obtained consistent (within uncertainties) coefficients of the relationship that scales the masses of the BH with redshift and luminosity. 

\section{Conclusions}

We have studied the Fe III$\lambda\lambda$2039-2113 emission line blend in 14 spectra of lensed quasars, in two well known AGN (NGC 5548 and NGC 7469), in the SDSS quasar composite spectrum and in 27 BOSS quasar spectra composites. This feature is relatively free of contamination from lines of other species and,  according to the impact of microlensing magnification on it, arises from an inner region of the BLR. The main results are:

1 - The Fe III$\lambda\lambda$2039-2113 feature appears systematically redshifted. In the high S/N ratio SDSS composite spectrum, this redshift is also consistently measured in the Fe III$\lambda$2419 line and the Fe III$\lambda\lambda$1970-2039 blend. 

2 - There is a correlation, though with a large scatter, between the observed redshift and the broadening of the Fe III$\lambda\lambda$2039-2113 lines. This dependence is expected in the case of virialized kinematics if the redshift is gravitational. The scatter may reflect the differences in geometry, kinematics and impact of non gravitational forces, among the quasars.

3 - In combination with microlensing based estimates of the Fe III UV emitting region size, the measured redshifts for gravitational lenses lead, under the gravitational redshift hypothesis, to values for the central BH mass, $\langle M_{BH}\rangle \simeq (0.9\pm 0.5)\times 10^9 M_\odot$, in good agreement with previous virial based estimates.

4 - We present a scaling relationship of mass with redshift and luminosity useful to measure the BH mass of one individual object from a single spectrum. This relationship can be formally derived from the Schwarzschild metric and is consistently calibrated using three different methods:  the broadening of the CIV lines of the BOSS composite spectra, the strength of gravitational microlensing in the Fe III UV lines, and the best fit to the available virial masses. The two first methods are completely independent and the estimated masses using any of them are in statistical agreement with virial masses over two orders of magnitude ($1\,\sigma$  scatter of 0.27 dex comparable to the intrinsic scatter of the virial masses).   

5 - {If the gravitational redshift hypothesis is correct,} the application of the scaling relationship to spectra of available quasar surveys will provide thousands of estimates of supermassive BH masses. Future mass estimates based on the Fe III redshift and reverberation mapping may become the primary calibrator for all BH mass measurement methods.

Although the good matching between the masses derived from the measured redshifts of the Fe III$\lambda\lambda$2039-2113 feature and the virial masses makes gravitational redshift a compelling explanation, the potential importance of the confirmation of this hypothesis is worthy of additional study. However this is not straightforward. Because of the large intrinsic uncertainties of the virial method applied to individual objects, a direct confirmation based on the comparison with virial masses, can be firmly established only from a large enough sample. In addition, as virial masses are not exempt of biases arising from the geometry of the emitters distribution or by the presence of non gravitational forces, this comparison will actually be two-way, testing both, the conditions of applicability of the virial theorem and the gravitational redshift hypothesis. For these reasons, the most convincing support likely will  be based on high S/N reverberation mapping studies of the Fe III$\lambda\lambda$2039-2113 blend in several objects, which can confirm the small size of the region emitting this spectral feature and provide an accurate R-L relationship for it.

}
\acknowledgements{We are grateful to the anonymous referee for a thorough revision and valuable suggestions. We thank the SDSS and BOSS surveys for kindly providing the data. This research was supported by the Spanish MINECO with the grants AYA2013-47744-C3-3-P and AYA2013-47744-C3-1-P. J.A.M. is also supported by the Generalitat Valenciana with the grant PROMETEO/2014/60. J.J.V. is supported by the project AYA2014-53506-P financed by the Spanish Ministerio de Econom\'\i a y Competividad and by the Fondo Europeo de Desarrollo Regional (FEDER), and by project FQM-108 financed by Junta de Andaluc\'\i a.  V.M. gratefully acknowledges support from FONDECYT through grant 1120741 and Centro de Astrof\'\i sica de Valpara\'\i so.}

%
%
%


\begin{figure}[h]
\includegraphics[scale=0.75]{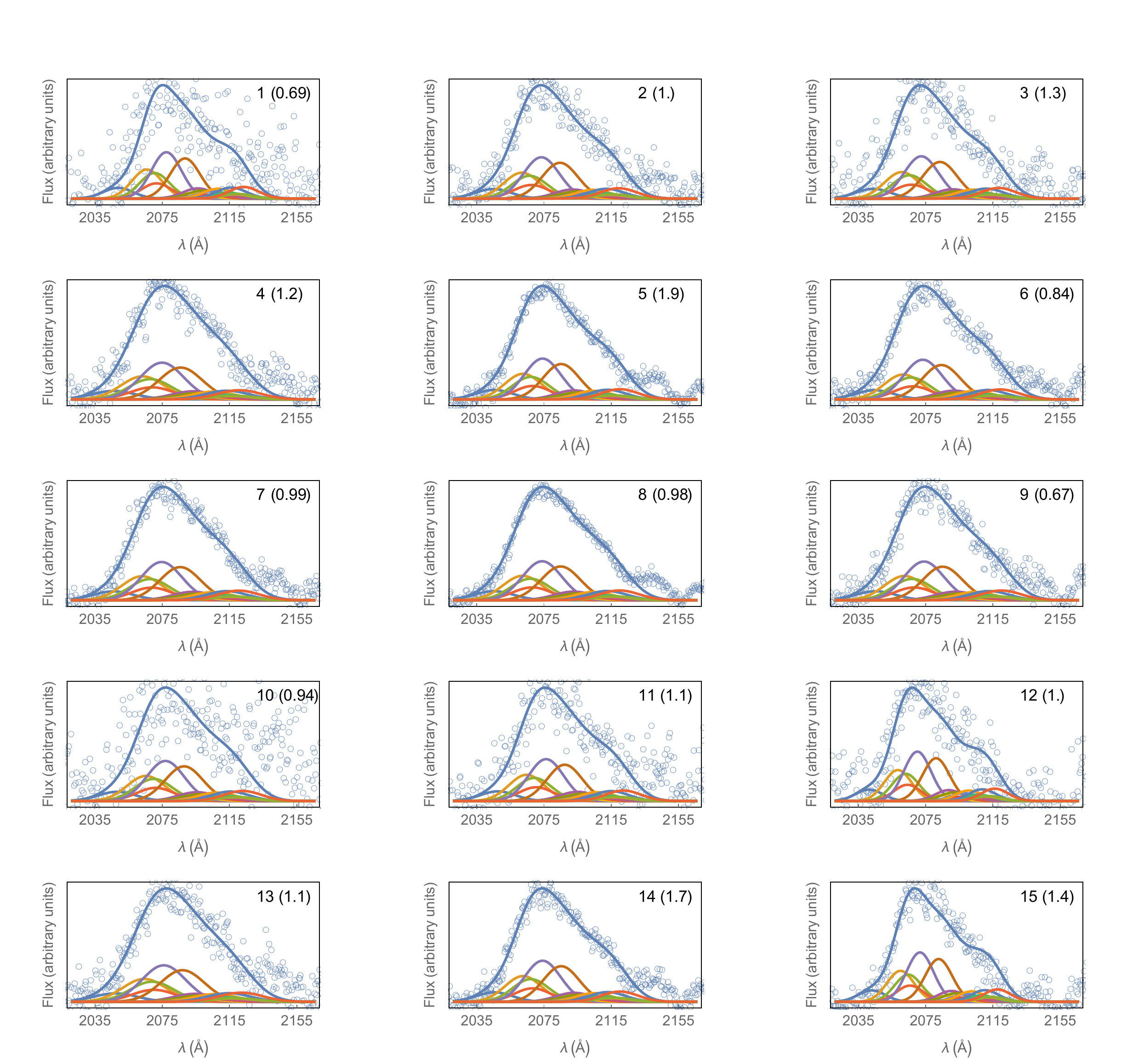}
\caption{Fits of {the} Vestergaard \& Wilkes (2001) template to the 27 composite spectra of the BOSS survey. Open circles correspond to the data, the blue line to the template and the other lines to the Gaussians representing each of the Fe III lines. The number of each composite is indicated and, in parentheses, the reduced chi-squared value, $\chi_{red}^2$ (see text). \label{grid}}
\end{figure}

\begin{figure}[h]
\includegraphics[scale=0.75]{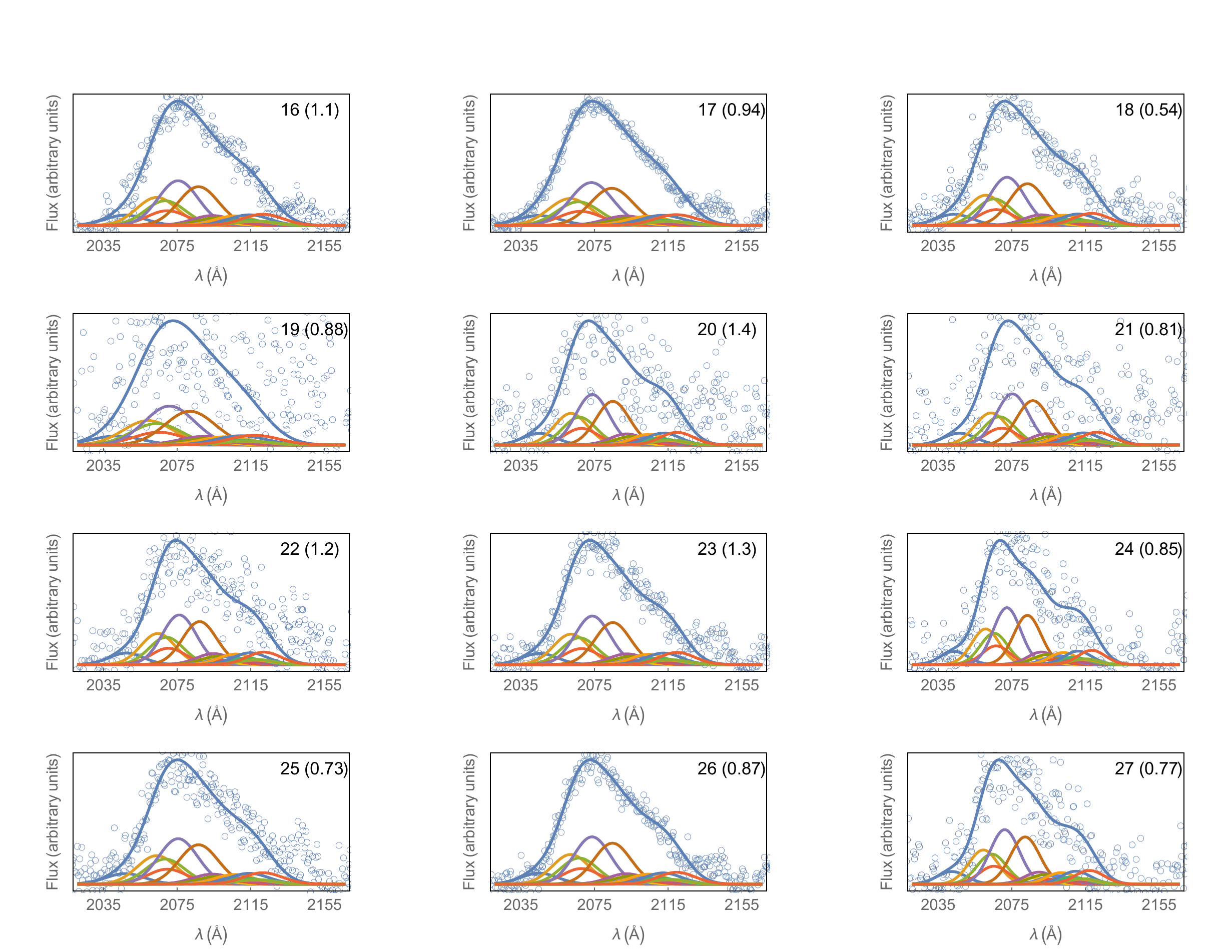}
\caption{Continuation of Figure \ref{grid}. \label{grid2}}
\end{figure}

\begin{figure}[h]
\plotone{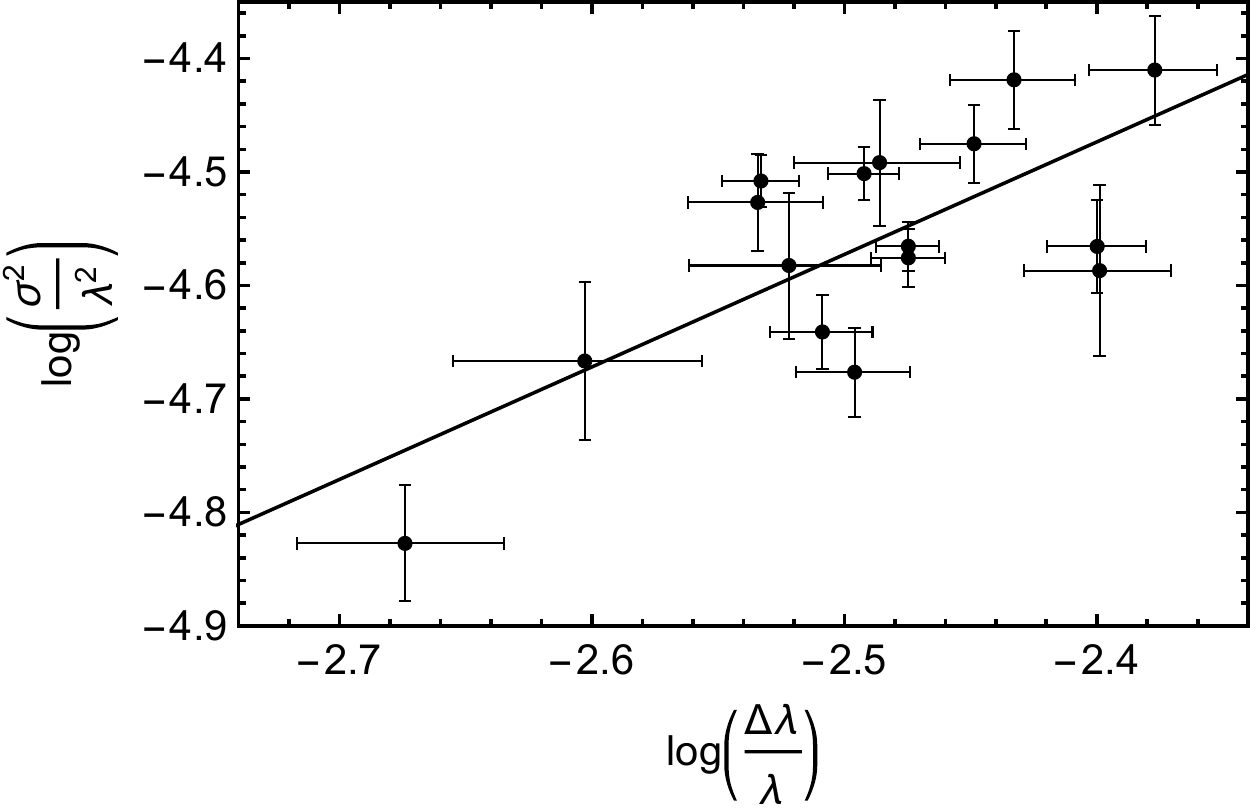}
\caption{Fe III$\lambda\lambda$2039-2113  width  squared, $\left({\sigma \over \lambda}\right)^2$, versus redshift, ${\Delta \lambda \over \lambda}$, obtained from the composite spectra of the BOSS survey with S/N $>$ 3. The straight line is the best fit to the data points (see text).\label{lambda_vs_sigma2}}
\end{figure}
%
%
%
%
\begin{figure}[h]
\plotone{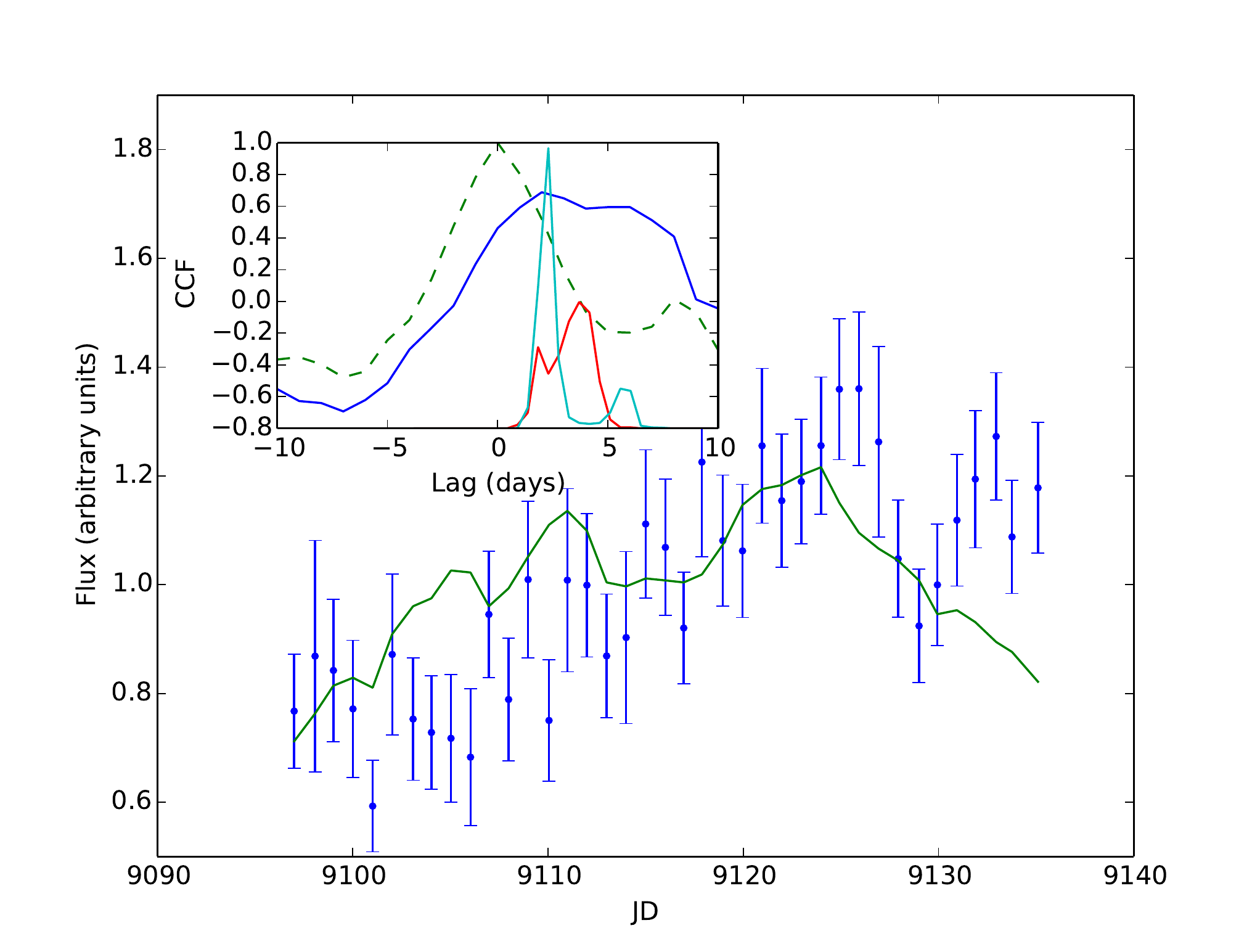}
\caption{Reverberation lag of Fe III$\lambda\lambda$2039-2113 for NGC 5548. Main plot: Fe III$\lambda\lambda$2039-2113 (blue dots) and UV continuum at 1790\AA\ (green line). Inset: CCF (Cross Correlation Function) in blue, Continuum ACF (Auto Correlation Function) in dashed green. CCCD (Cross Correlation Centroid Distribution) in red, CCPD (Cross Correlation Peak Distribution) in cyan. \label{NGC5548}}
\end{figure}

\begin{figure}[h]
\plotone{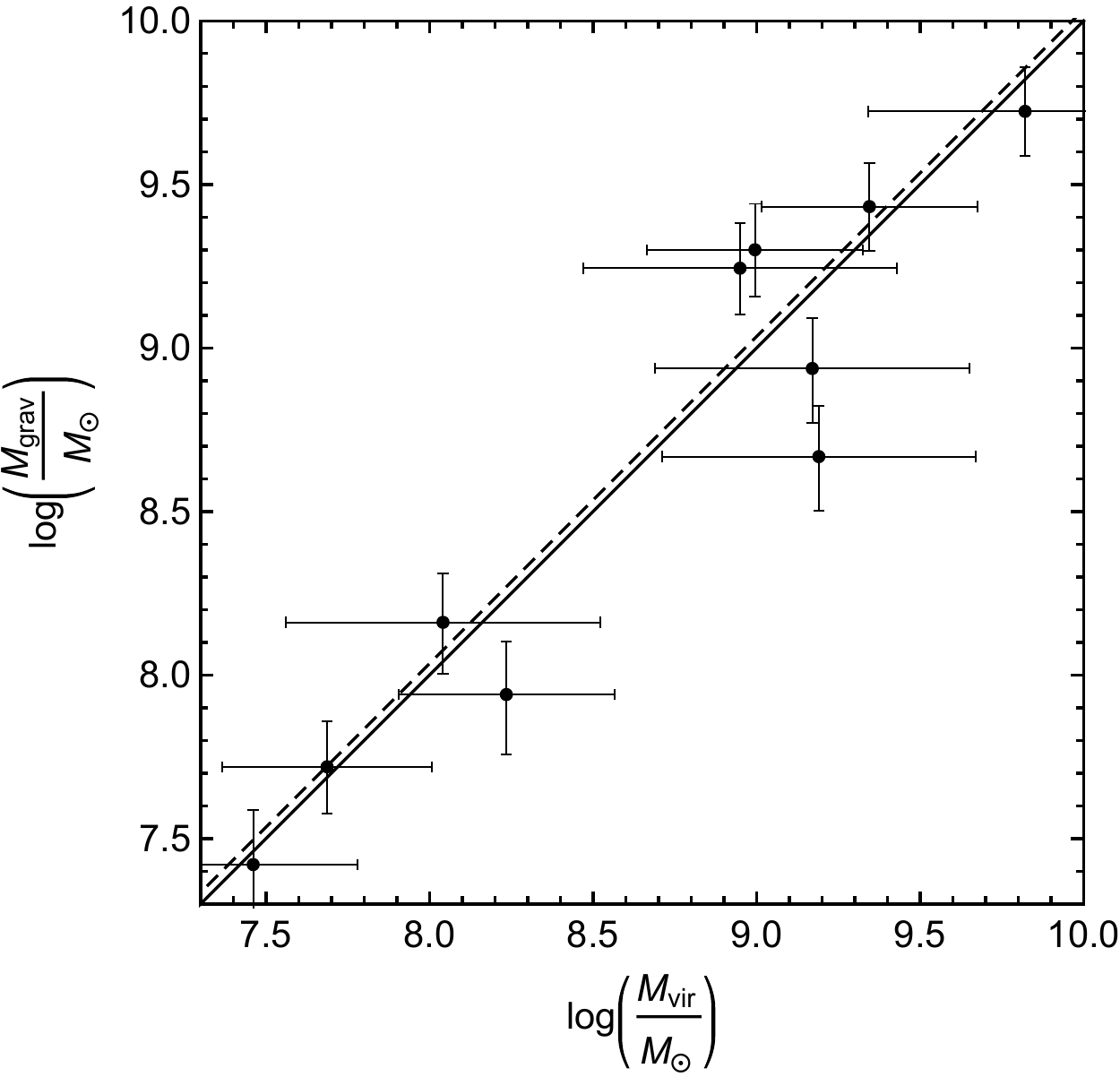}
\caption{Comparison between the virial and gravitational redshift based masses (calibration based on the widths of the CIV lines in the BOSS composite spectra). 
The solid line corresponds to $M^{BOSS}_{grav}=M_{vir}$. The dashed line corresponds to the best linear fit to the data with slope unity. The small separation between both lines indicates the good agreement between the BOSS based calibration and the calibration that would be obtained using the virial based mass estimates (see text). Errors in $M_{vir}$ are from Assef et al. (2011) or correspond to the dispersions of the virial relationships (Peng et al. 2006, Vestergaard \& Peterson 2006). Errors in $M^{BOSS}_{grav}$ include a (conservative) error of $\pm1.5\,$\AA\ in the gravitational redshift estimate, and 0.13 dex of intrinsic scatter in the R-L relationship (Peterson 2014). \label{mass_vs_mass_BOSS}}
\end{figure}

\begin{figure}[h]
\plotone{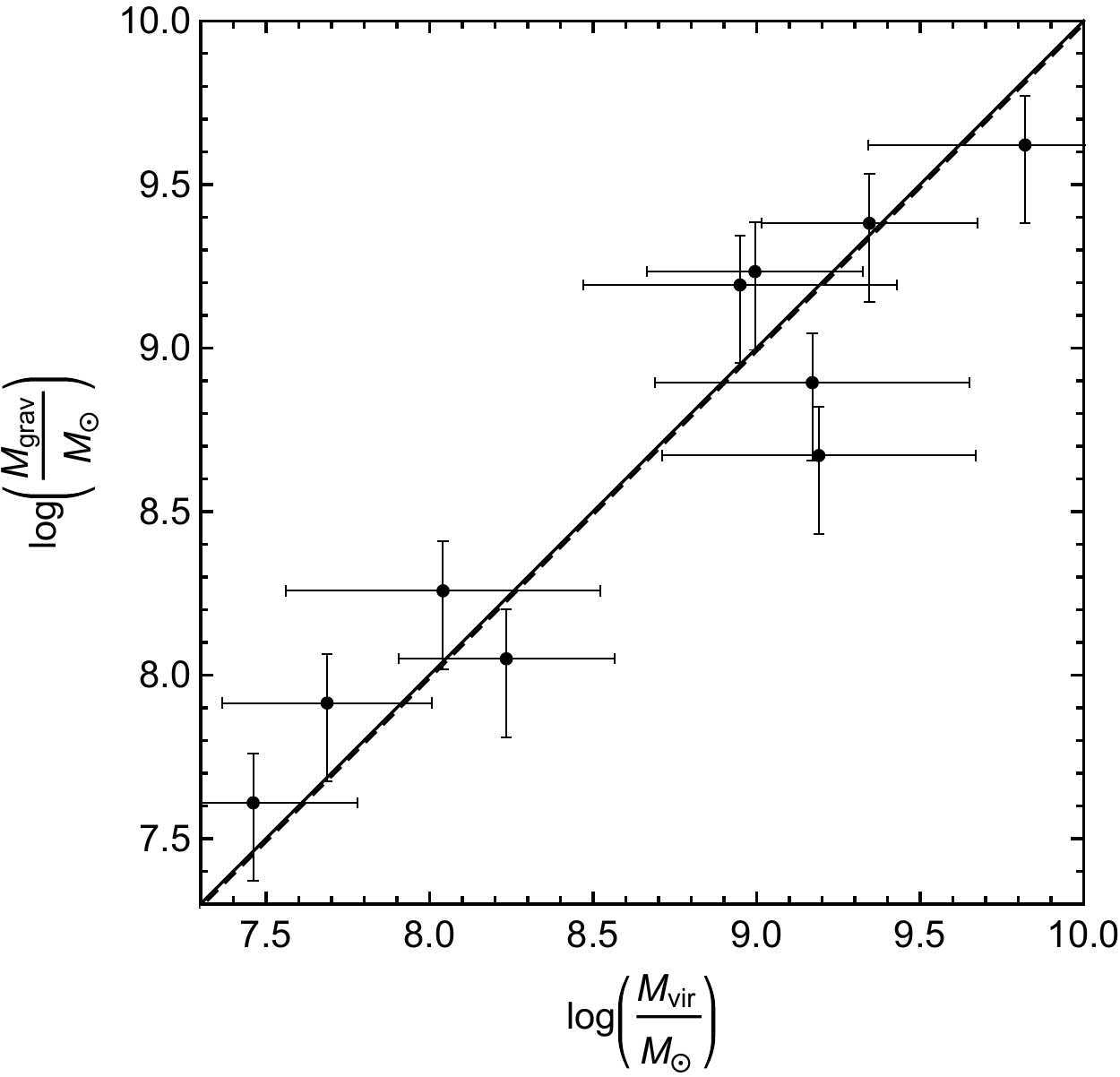}
\caption{Comparison between the virial and gravitational redshift based masses (calibrated using microlensing to determine a {reference size}). 
The continuous line corresponds to $M^{micro}_{grav}=M_{vir}$. The dashed line corresponds to the best linear fit to the data with slope unity. The very small {separation} between both lines indicates the excellent agreement between the microlensing based calibration and the calibration that would be obtained using the virial based mass estimates. Errors in $M_{vir}$ are from Assef et al. (2011) or correspond to the dispersions of the virial relationships (Peng et al. 2006, Vestergaard \& Peterson 2006). Errors in $M^{micro}_{grav}$ include the scatter in the estimate of the average gravitational redshift, the error in the microlensing estimate of the size, and 0.13 dex of intrinsic scatter in the R-L relationship (Peterson 2014). \label{mass_vs_mass_micro}}
\end{figure}

\begin{figure}[h]
\plotone{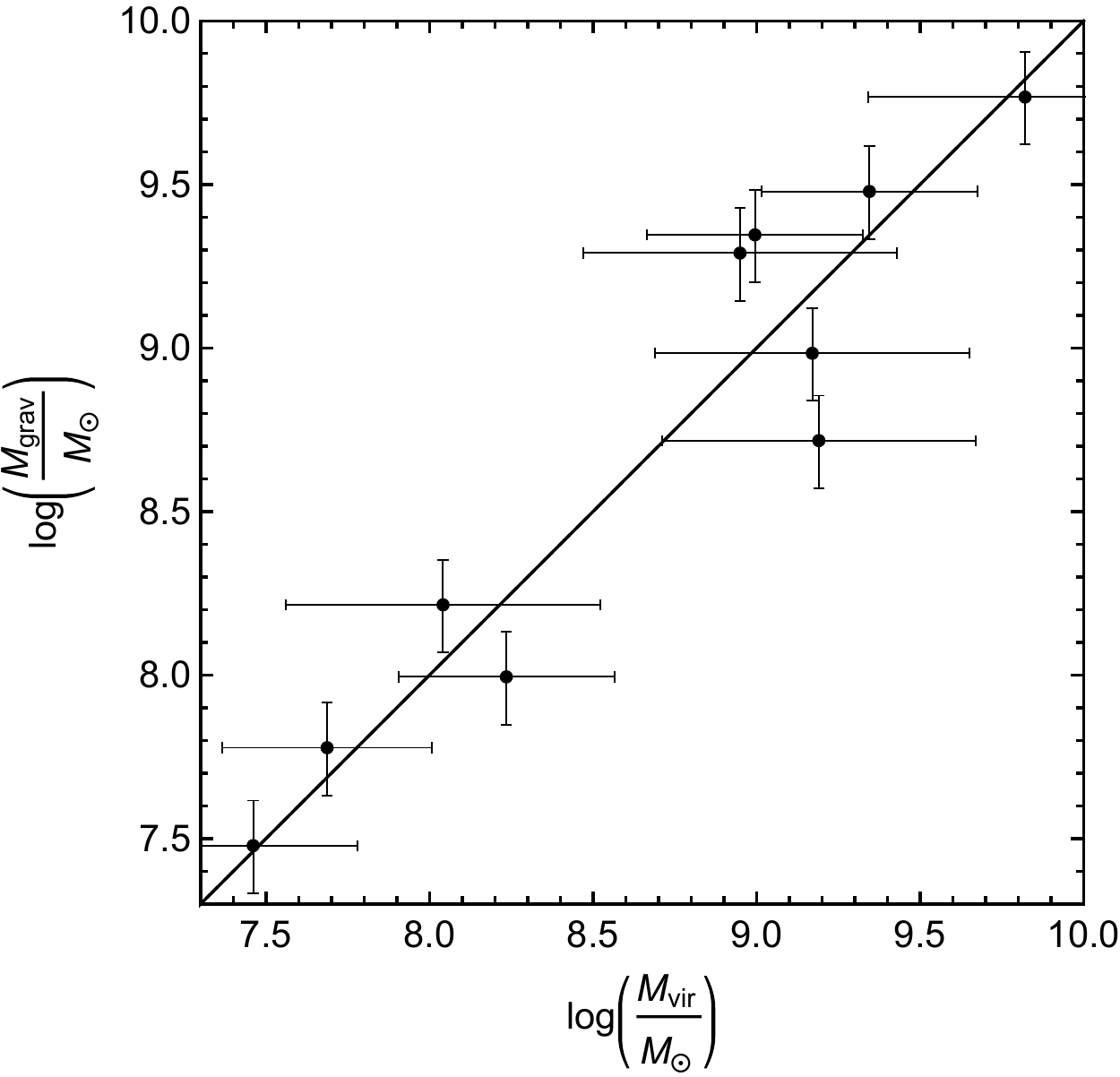}
\caption{Best fit of the mass scaling relationship based on the redshift to the virial masses leaving free the $R\propto L^b$ law. 
The continuous line corresponds to $M^{best\, fit}_{grav}=M_{vir}$. The best linear fit to the data with slope unity is indistinguishable from this line. Errors in $M^{best\, fit}_{vir}$ are from Assef et al. (2011) or correspond to the dispersions of the virial relationships (Peng et al. 2006, Vestergaard \& Peterson 2006). Errors in $M^{best\, fit}_{grav}$ include the error in the parameters of the fit and 0.13 dex of intrinsic scatter in the R-L relationship (Peterson 2014). \label{mass_vs_mass_bestfit}}
\end{figure}



\end{document}